\documentclass[%
twocolumn,letterpaper,
superscriptaddress,
showpacs,preprintnumbers,
nofootinbib,
nobibnotes,
bibnotes,
amsmath,amssymb,
aps,
prc,
floatfix,
]{revtex4-1}

\usepackage{graphicx}
\usepackage{dcolumn}
\usepackage{bm}
\usepackage{subfig}
\usepackage{siunitx}
\begin{document}


\title{Beam-energy dependence of identified two-particle angular correlations in Au+Au collisions at RHIC}

%
%
\affiliation{Abilene Christian University, Abilene, Texas   79699}
\affiliation{AGH University of Science and Technology, FPACS, Cracow 30-059, Poland}
\affiliation{Alikhanov Institute for Theoretical and Experimental Physics NRC "Kurchatov Institute", Moscow 117218, Russia}
\affiliation{Argonne National Laboratory, Argonne, Illinois 60439}
\affiliation{American University of Cairo, New Cairo 11835, New Cairo, Egypt}
\affiliation{Brookhaven National Laboratory, Upton, New York 11973}
\affiliation{University of California, Berkeley, California 94720}
\affiliation{University of California, Davis, California 95616}
\affiliation{University of California, Los Angeles, California 90095}
\affiliation{University of California, Riverside, California 92521}
\affiliation{Central China Normal University, Wuhan, Hubei 430079 }
\affiliation{University of Illinois at Chicago, Chicago, Illinois 60607}
\affiliation{Creighton University, Omaha, Nebraska 68178}
\affiliation{Czech Technical University in Prague, FNSPE, Prague 115 19, Czech Republic}
\affiliation{Technische Universit\"at Darmstadt, Darmstadt 64289, Germany}
\affiliation{ELTE E\"otv\"os Lor\'and University, Budapest, Hungary H-1117}
\affiliation{Frankfurt Institute for Advanced Studies FIAS, Frankfurt 60438, Germany}
\affiliation{Fudan University, Shanghai, 200433 }
\affiliation{University of Heidelberg, Heidelberg 69120, Germany }
\affiliation{University of Houston, Houston, Texas 77204}
\affiliation{Huzhou University, Huzhou, Zhejiang  313000}
\affiliation{Indian Institute of Science Education and Research (IISER), Berhampur 760010 , India}
\affiliation{Indian Institute of Science Education and Research (IISER) Tirupati, Tirupati 517507, India}
\affiliation{Indian Institute Technology, Patna, Bihar 801106, India}
\affiliation{Indiana University, Bloomington, Indiana 47408}
\affiliation{Institute of Modern Physics, Chinese Academy of Sciences, Lanzhou, Gansu 730000 }
\affiliation{University of Jammu, Jammu 180001, India}
\affiliation{Joint Institute for Nuclear Research, Dubna 141 980, Russia}
\affiliation{Kent State University, Kent, Ohio 44242}
\affiliation{University of Kentucky, Lexington, Kentucky 40506-0055}
\affiliation{Lawrence Berkeley National Laboratory, Berkeley, California 94720}
\affiliation{Lehigh University, Bethlehem, Pennsylvania 18015}
\affiliation{Max-Planck-Institut f\"ur Physik, Munich 80805, Germany}
\affiliation{Michigan State University, East Lansing, Michigan 48824}
\affiliation{National Research Nuclear University MEPhI, Moscow 115409, Russia}
\affiliation{National Institute of Science Education and Research, HBNI, Jatni 752050, India}
\affiliation{National Cheng Kung University, Tainan 70101 }
\affiliation{Nuclear Physics Institute of the CAS, Rez 250 68, Czech Republic}
\affiliation{Ohio State University, Columbus, Ohio 43210}
\affiliation{Panjab University, Chandigarh 160014, India}
\affiliation{Pennsylvania State University, University Park, Pennsylvania 16802}
\affiliation{NRC "Kurchatov Institute", Institute of High Energy Physics, Protvino 142281, Russia}
\affiliation{Purdue University, West Lafayette, Indiana 47907}
\affiliation{Rice University, Houston, Texas 77251}
\affiliation{Rutgers University, Piscataway, New Jersey 08854}
\affiliation{Universidade de S\~ao Paulo, S\~ao Paulo, Brazil 05314-970}
\affiliation{University of Science and Technology of China, Hefei, Anhui 230026}
\affiliation{Shandong University, Qingdao, Shandong 266237}
\affiliation{Shanghai Institute of Applied Physics, Chinese Academy of Sciences, Shanghai 201800}
\affiliation{Southern Connecticut State University, New Haven, Connecticut 06515}
\affiliation{State University of New York, Stony Brook, New York 11794}
\affiliation{Temple University, Philadelphia, Pennsylvania 19122}
\affiliation{Texas A\&M University, College Station, Texas 77843}
\affiliation{University of Texas, Austin, Texas 78712}
\affiliation{Tsinghua University, Beijing 100084}
\affiliation{University of Tsukuba, Tsukuba, Ibaraki 305-8571, Japan}
\affiliation{United States Naval Academy, Annapolis, Maryland 21402}
\affiliation{Valparaiso University, Valparaiso, Indiana 46383}
\affiliation{Variable Energy Cyclotron Centre, Kolkata 700064, India}
\affiliation{Warsaw University of Technology, Warsaw 00-661, Poland}
\affiliation{Wayne State University, Detroit, Michigan 48201}
\affiliation{Yale University, New Haven, Connecticut 06520}

\author{J.~Adam}\affiliation{Brookhaven National Laboratory, Upton, New York 11973}
\author{L.~Adamczyk}\affiliation{AGH University of Science and Technology, FPACS, Cracow 30-059, Poland}
\author{J.~R.~Adams}\affiliation{Ohio State University, Columbus, Ohio 43210}
\author{J.~K.~Adkins}\affiliation{University of Kentucky, Lexington, Kentucky 40506-0055}
\author{G.~Agakishiev}\affiliation{Joint Institute for Nuclear Research, Dubna 141 980, Russia}
\author{M.~M.~Aggarwal}\affiliation{Panjab University, Chandigarh 160014, India}
\author{Z.~Ahammed}\affiliation{Variable Energy Cyclotron Centre, Kolkata 700064, India}
\author{I.~Alekseev}\affiliation{Alikhanov Institute for Theoretical and Experimental Physics NRC "Kurchatov Institute", Moscow 117218, Russia}\affiliation{National Research Nuclear University MEPhI, Moscow 115409, Russia}
\author{D.~M.~Anderson}\affiliation{Texas A\&M University, College Station, Texas 77843}
\author{A.~Aparin}\affiliation{Joint Institute for Nuclear Research, Dubna 141 980, Russia}
% ...

\collaboration{STAR Collaboration}\noaffiliation

\date{\today}

\begin{abstract}
The two-particle angular correlation functions, $R_2$, of pions, kaons, and
protons in Au+Au collisions at $\sqrt{s_{NN}}=$ 7.7, 11.5, 14.5,
19.6, 27, 39, 62.4, and 200 GeV were measured by the STAR
experiment at RHIC. These correlations were measured for both
like-sign and unlike-sign charge combinations and versus the
centrality. The correlations of pions and kaons show the expected
near-side ({\it i.e.}, at small relative angles) peak resulting from short-range 
mechanisms. The amplitudes of these short-range correlations decrease with
increasing beam energy. However, the proton correlation functions
exhibit strong anticorrelations in the near-side region.  This behavior is observed for
the first time in an A+A collision system. The observed
anticorrelation is $p_{T}$-independent and decreases with
increasing beam energy and centrality. The experimental results
are also compared to the Monte Carlo models UrQMD, Hijing, and
AMPT.
\end{abstract}


\maketitle

\section{Introduction}

The study of single-particle observables provides information on the bulk
properties of the hot nuclear systems formed in relativistic heavy-ion
collisions. A more differential view, first employed to understand the
systems produced at the ISR in the 1970's \cite{Foa1975,DeWolf:1995nyp,Eggert1974, UA5},
involves the use of two-particle correlators. Here, one measures the rates
for all pairs of particles in single events versus kinematic observables in
two dimensions, {\it e.g.}, the relative rapidity and azimuthal angle,
$(\Delta y,\Delta\varphi)$, of the two particles in each pair. These
distributions can then be normalized by the distributions formed once the
intraevent correlations have been explicitly broken. This normalization
also removes any contributions to the correlators from all single-particle
inefficiencies in the experimental measurement. The resulting ratio, called
$R_2$, then depicts excesses or deficits with respect to unity that
directly indicate correlations or anticorrelations, respectively. Parton
fragmentation, resonance decays, and femtoscopic correlations, typically
referred to as ``short-range" correlations, are localized to a narrow
region near $(\Delta y,\Delta\varphi)\sim0$ \cite{Jet2017, Femtoscopy}.
Other phenomena are longer range, such as elliptic flow, which appear as a
cosine function of the relative azimuthal angle \cite{ATLAS2012}.  Global
momentum conservation can result in a back-to-back correlation between the
produced particles, which is reflected as a negative cosine function of
$\Delta\varphi$ \cite{Borghini2002, Borghini2000, ATLAS2012}. Non-zero
integrals of the two-particle correlation functions result in multiplicity
distributions with variances that are not equal to the mean values, as
would be expected for purely Poisson fluctuations. As the variance of the
multiplicity distributions goes like the square of the correlation length
\cite{Hatta2003wn}, 
the two-particle correlation functions thus provide a
more differential view of effects which may potentially result from
the proximity of a critical point \cite{Hatta2003wn,Stephanov:1998dy,Stephanov:1999zu,Stephanov:2001zj,Antoniou:2001rx,Koch:2001zn,Pruneau:2002yf}.
Such a critical point would be expected to mark the end of the first-order phase transition line
separating hadronic and partonic matter. The expected critical point signal is thus a nonmonotonic dependence of the fluctuations and correlations on the beam energy. Therefore, multiparticle correlations, and their integrals the fluctuations,
deserve careful study.

In this paper, the two-particle correlations are studied for like-sign and
unlike-sign identified pions, kaons, and protons in Au+Au collisions
measured by the STAR experiment during the Beam Energy Scan (BES) program
at RHIC. The angular correlation functions are presented at eight
different beam energies ranging from 7.7 to 200 GeV and at three selected centralities, 
the most central 0\%-5\%, 30\%-40\%, and peripheral
60\%-70\%. Two ranges of low and high transverse momentum are also compared. 
The study of the different particle species pairs allows one to compare the meson ($\pi$
and K) and baryon (p) correlations. The beam energy dependence spans
nearly baryon-free matter at the highest energy to increasingly baryon-doped matter 
as the beam energy is decreased. The
experimental results are also compared to those from the models UrQMD
\cite{Bass1998}, Hijing \cite{Wang1991}, and AMPT \cite{Lin2005}, each of
which produces events based on different theoretical approaches.

This paper is organized as follows: the STAR detector and other
experimental details are described in Section \ref{sec:experimental}; the
two-particle angular correlation function results are presented in Section
\ref{sec:results}. Finally, the summary and conclusions are presented in
Section \ref{sec:summary}.

\section{Experimental details\label{sec:experimental}}

The Solenoidal Tracker at RHIC (STAR) is an azimuthally-symmetric and wide
acceptance detector. The key subdetectors used here include the Time
Projection Chamber (TPC) \cite{STARtpc}, which performs the track and primary
vertex reconstruction, as well as particle identification at low momentum,
and the Time-of-Flight system (TOF) \cite{STARtof}, which provides particle
identification information at higher momentum. A solenoidal magnet aligned
with the beam axis provides a uniform magnetic field of 0.5 T for charged
particle momentum analysis \cite{Green1993}.

The data studied here were collected in the years 2010, 2011, and 2014, and
include the eight beam energies of $\sqrt{s_{NN}}=$ 7.7, 11.5, 14.5, 19.6,
27, 39, 62.4, and 200 GeV. These data were collected with a minimum bias
trigger based on the information from the Vertex Position Detector (VPD)
\cite{STARvpd}, Beam-Beam Counters (BBC) and Zero Degree Calorimeter (ZDC)
detectors \cite{Judd}. The raw event totals and the year of data collection
are shown in Table \ref{table1}.  

\begin{table}[h!]
\caption{\small The number of events and year the data was taken versus the beam energy. }
\label{table1}
\begin{ruledtabular}
\begin{tabular}{c c c } 
$\sqrt{s_{NN}}$ & Year & $N_{\rm events}$ \\ 
      (GeV)          &          & (million)   \\
 \hline
 7.7   & 2010 & 3.2 \\ 
 11.5 & 2010 & 11.4  \\ 
 14.5 & 2014 & 15.9  \\ 
 19.6 & 2011 & 17.1  \\ 
 27    & 2011 & 31.3  \\ 
 39    & 2010 & 36.8  \\ 
 62.4 & 2010 & 39  \\ 
 200  & 2010 & 59.3  \\ 
\end{tabular}
\end{ruledtabular}
\end{table}

The collision vertex, determined using all charged tracks in each event, 
was required to be within $\pm$30 cm of the center of STAR
along the beam direction at all eight beam energies. 
Pseudocorrelations caused by the event by event variation of the location of the
primary vertex along the beam pipe, Z$_{\rm vtx}$, were removed by performing the analyses
in 30 bins of Z$_{\rm vtx}$, each 2 cm wide. A weighted average of the correlation functions
over these bins was then constructed, eliminating these pseudocorrelations \cite{Tarini}.

For the pion or kaon correlations, the centrality of the collisions was
determined using the charged particle multiplicity distributions with
pseudorapidities, $\eta$, within the range $0.5<|\eta|<1$ and a Monte Carlo Glauber simulation as described, {\it
e.g.}, in Ref. \cite{Miller2007glauber}. For the proton correlations, the centrality was
determined using the measured multiplicity of tracks, excluding protons,
with $|\eta|<1$. These same centrality definitions were used in the
STAR papers on the multiplicity cumulants \cite{STARnetp, STARnetq,
STARnetk}. To avoid artifacts in the observables of interest caused by 
the above multiplicity binning on pseudorapidity, the correlation
functions were studied only for pseudorapidities within the range
$|\eta|<0.5$. 

The raw events collected by STAR were then pruned of data-taking runs in
which the average values of a number of observables deviated by two
standard deviations from their values over all events. Examples of the
variables studied are the mean values of several different track or hit
multiplicities, or the average values of track-based quantities such as the
transverse momentum or azimuthal angle. About thirty such variables were
studied in each run, and the most sensitive to ``bad runs" were generally
the number of primary reconstructed tracks per event, the number of tracks per event
that matched to TOF hits, the east-west asymmetry in the track pseudorapidity,
and the averages of the track transverse and total momentum.
Once the bad runs were removed, multiple selection
criteria on pairs of global observables were applied to remove bad events
in good runs. These selection criteria were effective at removing
collisions of gold nuclei with beam line materials (most importantly at the
lowest beam energies) and collision pile-up in time in the TPC (most
importantly at the highest beam energies). The tracks used in the
correlations analyses were subject to quality cuts on the distance of
closest approach to the primary vertex (maximum 2 cm), the number of TPC
space points assigned to each track (minimum 18), and the ratio of assigned
to total possible space points (minimum 52\%). 

\begin{table}[h!]
\caption{\small The kinematic acceptance in rapidity and transverse momentum
for pions, kaons, and protons in this analysis. }
\label{tablecuts}  
\begin{ruledtabular}
\begin{tabular}{c l l } 
 $\pi^{\pm}$ 		& 0.2$<$$p_{T}$$<$2.0 GeV/c	&  $|y|$$<$0.42   \\
 K$^{\pm}$  		& 0.2$<$$p_{T}$$<$1.6 GeV/c	&  $|y|$$<$0.40   \\
 p,$\bar{\mathrm p}$	& 0.4$<$$p_{T}$$<$2.0 GeV/c	&  $|y|$$<$0.60   \\
 \end{tabular}
\end{ruledtabular}
\end{table}

The correlation functions were measured using like-sign (LS) and
unlike-sign (US) pairs of pions, kaons, and protons, separately. The
kinematic acceptance used for the different particle species is
shown in Table \ref{tablecuts}. 
To identify the particles, the ionization energy loss, $dE/dx$, measured by the TPC and the time
of flight measured by the TOF detector was used. The $dE/dx$ selection was done
within two standard deviations of each particle's peak in the normalized
ionization energy loss distributions. The TOF efficiency
per TPC track is $\sim$60-70\%. If the TOF information was available
for a given TPC track, a cut was placed on the mass obtained from the track
momentum and speed. If a particular track did not have TOF information,
additional exclusionary $dE/dx$ cuts on nearby particle species were 
applied at low momenta. 

By definition, the correlation functions, $R_2$, are insensitive to
single-particle experimental inefficiencies caused, for example, by gaps in
the detector. However, ``track crossing," a true two-particle
inefficiency, remains. The track reconstruction algorithm used in STAR
does not share space points between two nearby tracks. The imposition of
even minimal quality cuts on the number of space points assigned to a
reconstructed track thus causes one of the tracks in the pair to have
fewer space points and thus a slightly lower efficiency. This relative
inefficiency for finding a track because of the existence of another
nearby creates a ``near-side," $(\Delta y,\Delta\varphi)\sim0$, hole
in the correlation functions. This was avoided in the present analysis by
$p_{T}$-ordering the particles in each pair to constrain the track
crossing inefficiency to a smaller region, then reflecting the unaffected
bins across $\Delta\varphi=\Delta y=0$ to form the correlation
functions devoid of track crossing \cite{Tarini}. 
The affected regions for each particle species are summarized in
Table \ref{tablecross}. Additional systematic uncertainties result from the
specific treatment of the track crossing inefficiency and these can be
seen in the results below for the few bins very close to $\Delta y=0$.

%
%
\begin{table}[h!]
\caption{\small The kinematic regions affected by the track crossing inefficiency and
subsequent correction for each particle species. }
\label{tablecross}  
\begin{ruledtabular}
\begin{tabular}{c c l } 
 $\pi^{\pm}$		& $|\Delta y|<0.09$	& LS: ~~$-5^{\circ}<|\Delta\varphi|<35^{\circ}$	\\
				&  				& US: $-85^{\circ}<|\Delta\varphi|<-5^{\circ}$	\\
 K$^{\pm}$ 		& $|\Delta y|<0.12$	& LS: ~~$-5^{\circ}<|\Delta\varphi|<35^{\circ}$		   \\
				&  				& US: $-85^{\circ}<|\Delta\varphi|<-5^{\circ}$		\\
 p,$\bar{\mathrm p}$	& $|\Delta y|<0.20$	& LS: ~~$-5^{\circ}<|\Delta\varphi|<25^{\circ}$		   \\
				&  				& US: $-35^{\circ}<|\Delta\varphi|<-5^{\circ}$		\\
 \end{tabular}
\end{ruledtabular}
\end{table}

\subsection{Two-particle angular correlation functions}

The correlation function is defined as the ratio of the two-particle density distributions
and the product, or convolution, of the single-particle densities. This division
normalizes the correlations as ``per pair," and makes them insensitive to
single-particle reconstruction and acceptance inefficiencies \cite{Foa1975,Pruneau:2002yf,Ravan:2013lwa}. The normalized
``angular correlations," $R_{2}$, are formed as a function of the relative rapidity and 
azimuthal angle of the two particles in the pair, $(\Delta y,\Delta\varphi)$, 
and are given by \cite{Foa1975,DeWolf:1995nyp,Pruneau:2002yf,Kittel:2005qy,Kittel:2001gt,Ravan:2013lwa}:
\begin{equation}\label{formula1}
R_{2}(\Delta y,\Delta\varphi) =\frac{\rho_{2}(\Delta y,\Delta\varphi)}{\rho_{1}(y_{1},\varphi_{1})\rho_{1}(y_{2},\varphi_{2})}-1 ,
\end{equation}
where $\Delta y=y_{1}-y_{2}$, $\Delta\varphi=\varphi_{1}-\varphi_{2}$, and  
$\rho_{2}(\Delta y,\Delta\varphi)$ and $\rho_{1}(y,\varphi)$ are the two-particle and single-particle multiplicity density distributions,
respectively, normalized to the number of events. 

The numerator of the correlation functions for particles is calculated using all pairs in each
event except self-pairs. 
Several methods are available to calculate the denominator. 
These include pulling particles of interest
from two different but similar events, which is called ``mixing," and convolution.
In convolution, a single-particle spectrum versus $(y,\varphi, p_{T})$
is folded with itself in six nested loops to produce the denominator 
versus the pair $(\Delta y,\Delta\varphi)$. This six-dimension convolution allows
one to impose the same cut (see previous section) in the denominator
as was used in the numerator to remove the two-particle inefficiency from
track crossing. The results from the two methods to form the denominator,
mixing and convolution, were found to be in excellent agreement. 

The amplitudes of such $R_2$ correlation functions often decrease with increasing
beam energy and/or centrality as a result of the increasing number of particle-emitting 
sources for higher-energy (and/or more central) collisions. One may thus consider scaling
the correlators with some multiplicity such as the number of participants or binary collisions
to account for such dilution. The correlators shown here do not include such an additional scaling.

\begin{figure*}[htb] 
 \subfloat[Like-sign pions]{\includegraphics[trim= 0cm 0cm 0cm 0cm,clip,scale=0.45]{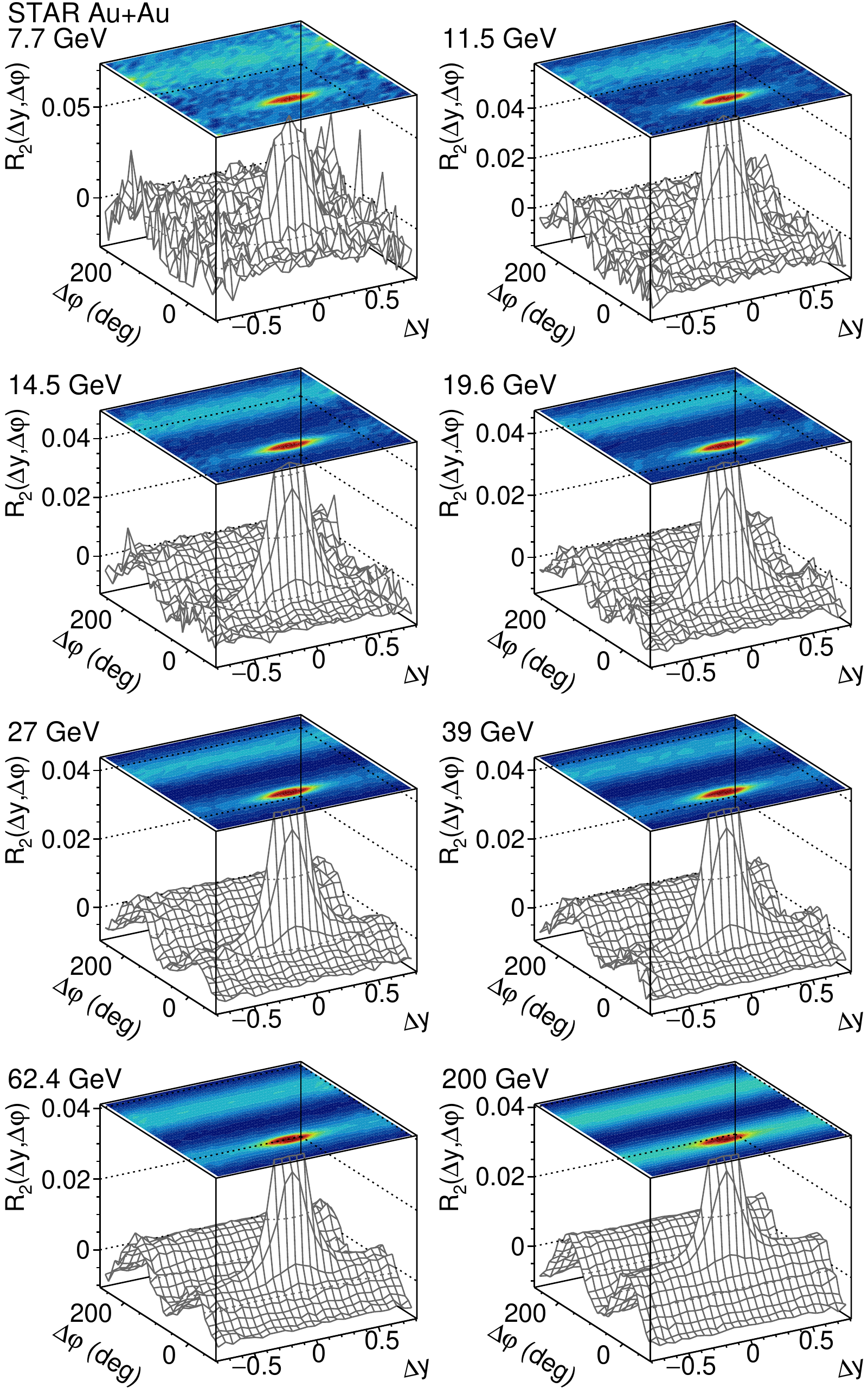} }
    \qquad \qquad
    \subfloat[Unlike-sign pions]{\includegraphics[trim= 0cm 0cm 0cm 0cm,clip,scale=0.45]{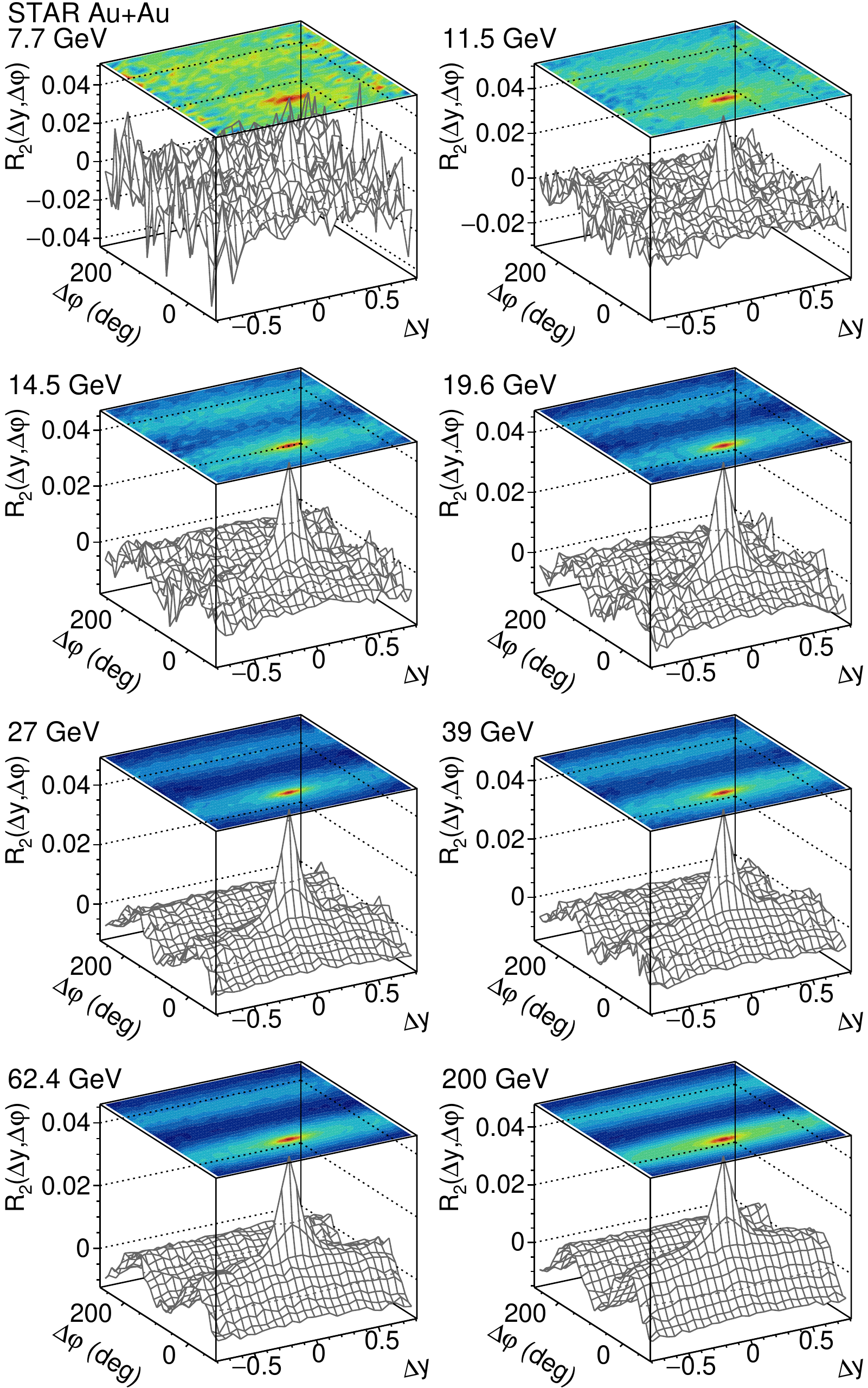} }
    \caption{\small Angular correlation function $R_{2}(\Delta y,\Delta\varphi)$ of like-sign (left) and unlike-sign (right) 
pions in Au+Au collisions at mid centrality 30\%-40\% and $0.2<p_{T}<2.0$ GeV/$c$ in 
different beam energies from 7.7 GeV (top left) to 200 GeV (bottom right). }
\label{plot_R2_pi}
\end{figure*}

In the present analysis, the numerator and denominator of the correlation functions were further
normalized to the event-averaged number of pairs \cite{Foa1975} via,
\begin{equation}\label{formula2}
R_{2}(\Delta y,\Delta\varphi) =\frac{\langle n\rangle^{2}}{\langle n(n-1)\rangle}\frac{ \rho_{2}(\Delta y,\Delta\varphi)}{\rho_{1}(y_{1},\varphi_{1})\rho_{1}(y_{2},\varphi_{2})} -1,
\end{equation}  
where $n$ is the event-by-event multiplicity of the indistinguishable particle of interest
in a given centrality and Z$_{\rm vtx}$ bin. If the particles in the pair are 
distinguishable, this prefactor becomes 
$\langle n_1 \rangle \langle n_2 \rangle / \langle n_1 n_2 \rangle $, where $n_1$ and $n_2$
are the event-wise multiplicities of the distinguishable particles of interest.  
This normalization removes
purely mathematical finite-multiplicity offsets to the correlation
functions and thus ensures that the values of $R_2$ are identically zero
in the absence of any two-particle (anti)correlations even at low
multiplicities of the particle of interest in each event. 

\subsection{Systematic uncertainty}

To estimate the systematic uncertainties, the track selection 
and particle identification criteria were modified within reasonable ranges,
and the full analysis was repeated for each cuts set. The systematic
uncertainties for the track selection and particle identification were
separately studied. The standard deviation of the results when using the
default cut was calculated for each set and the systematic uncertainty
was determined as the root of the quadratic sum of the different
systematic sources. 

\begin{figure*}[htb] 
    \subfloat[Like-sign protons]{\includegraphics[trim= 0cm 0cm 0cm 0cm,clip,scale=0.45]{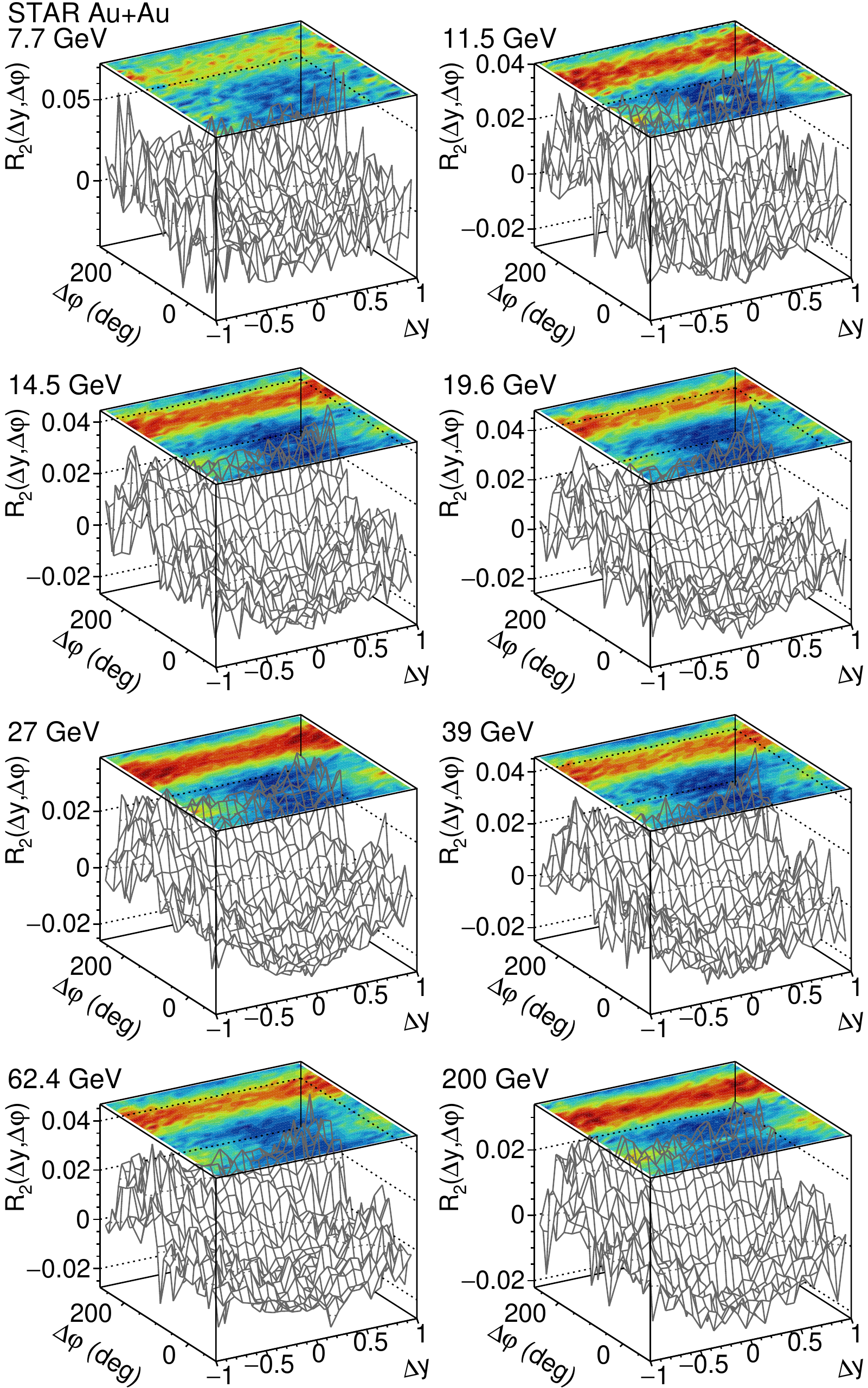} }
    \qquad \qquad
    \subfloat[Unlike-sign protons]{\includegraphics[trim= 0cm 0cm 0cm 0cm,clip,scale=0.45]{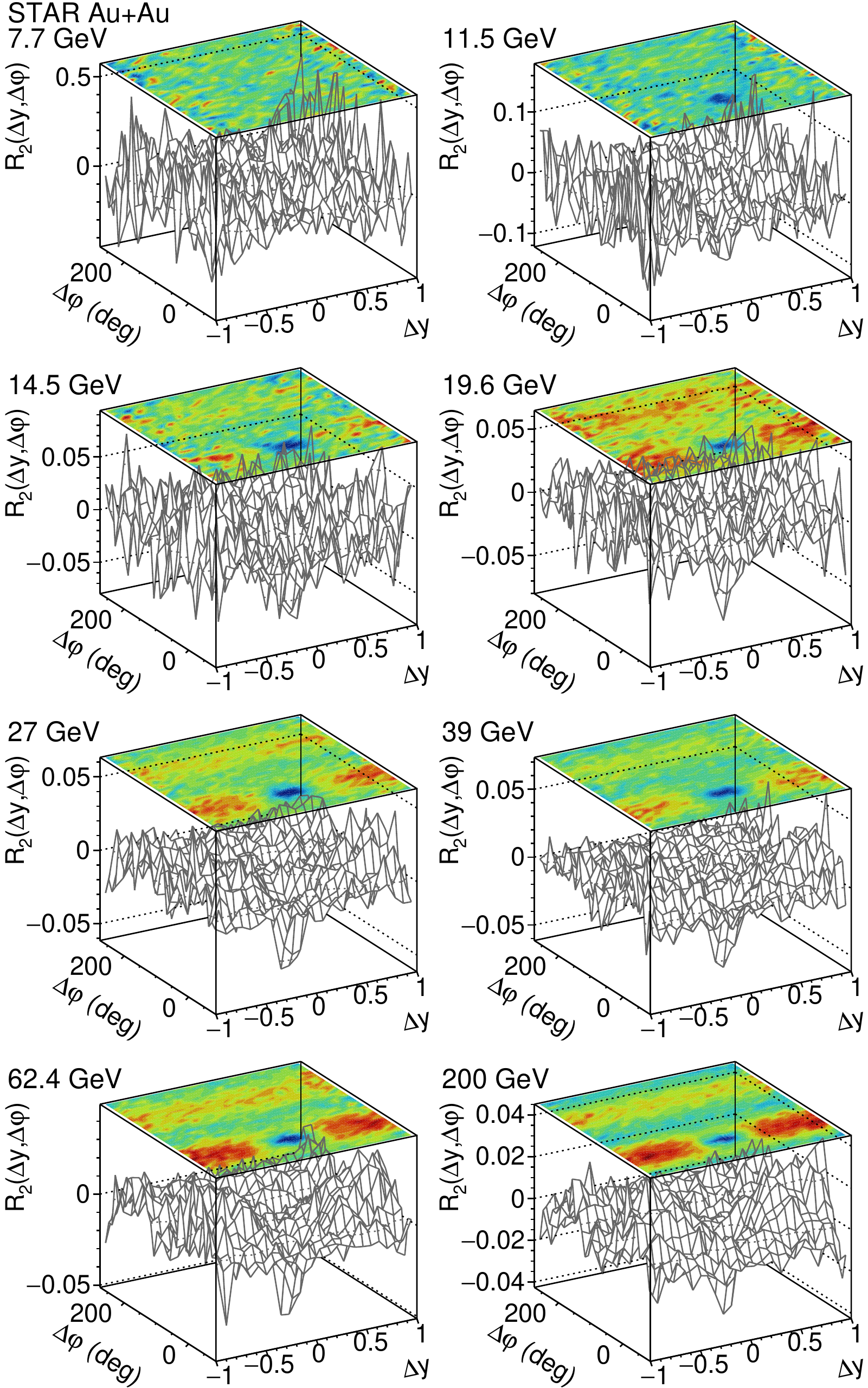} }
    \caption{\small Angular correlation function $R_{2}(\Delta y,\Delta\varphi)$ of like-sign (left) and unlike-sign 
(right) protons in Au+Au collisions at mid centrality 30\%-40\% and $0.4<p_{T}<2.0$ 
GeV/$c$ in different beam energies from 7.7 GeV (top left) to 200 GeV (bottom right). Due to the 
large statistical fluctuations in large $\Delta y$ bins, the plots are presented in the range 
of $|\Delta y|\leq1$.  }
\label{plot_R2_p}    
\end{figure*}

The main source of systematic uncertainty for the pion results was the
cut on the distance of closest approach to the primary vertex. For the
kaon and proton results, the particle identification cuts resulted in the
largest contributions in the systematic uncertainties. The absolute uncertainties
of the main systematic source averaged over $\Delta y$ at 62.4 GeV, 30\%-40\% centrality, were
found to be \num{0.1e-3} for like-sign and unlike-sign pions, \num{0.3e-3} for
like-sign kaons and protons, and lower than \num{0.5e-3} for unlike-sign kaons
and protons. The systematic uncertainties at 14.5 GeV, and 30\%-40\%
centrality, are similar, although they increase to \num{0.8e-3} for like-sign
kaons, and \num{0.1e-2} for unlike-sign kaons and protons. The final source of
systematic uncertainty results from the necessary correction for 
the track crossing pair inefficiency. This contribution can be larger than
the other systematics but only for the few bins near $\Delta y=0$, as will
be seen in the results presented below. 

\section{Results\label{sec:results}}

\begin{figure}[hb] 
    \includegraphics[trim= 0cm 0cm 0cm 0cm, scale=0.41]{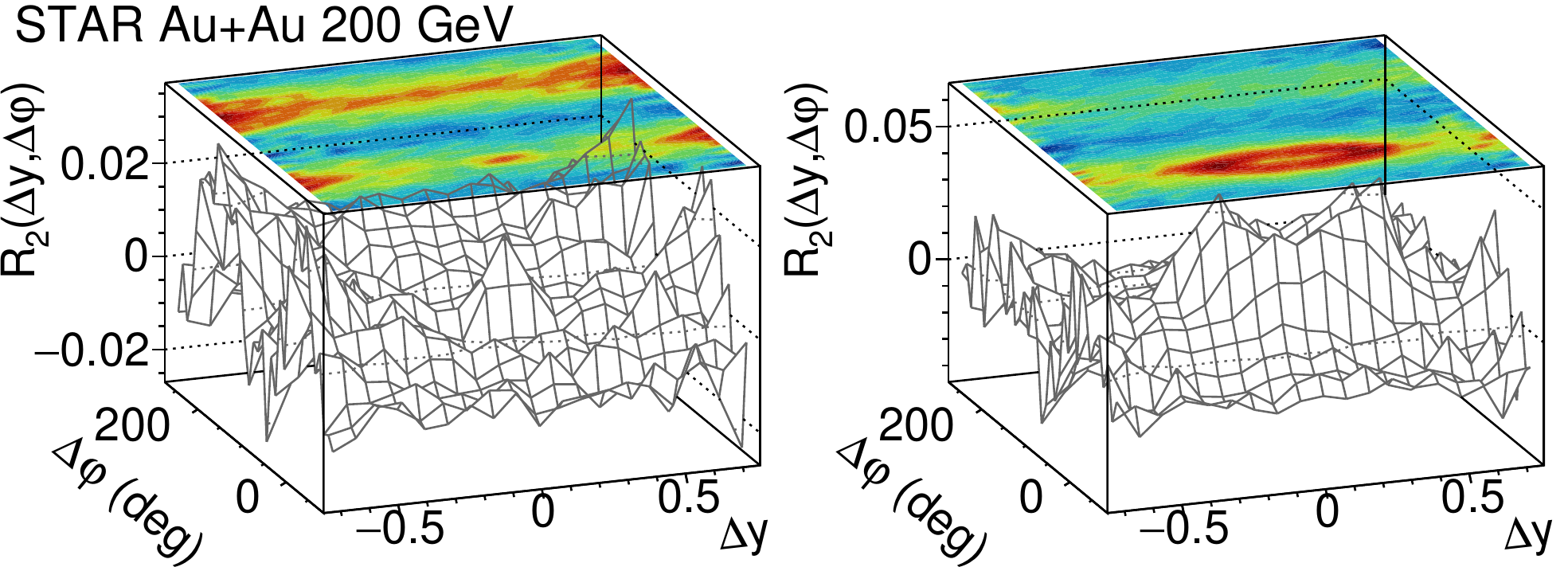}
    \caption{\small Angular correlation function $R_{2}(\Delta y,\Delta\varphi)$ of like-sign (left) and unlike-sign 
(right) kaons in Au+Au collisions at 200 GeV, mid centrality 30\%-40\% and $0.2<p_{T}<1.6$ GeV/$c$.} 
\label{plot_R2_K}
\end{figure}

The angular correlation functions for like-sign and unlike-sign
identified $\pi$ mesons and protons are shown in Fig. \ref{plot_R2_pi}
and \ref{plot_R2_p}, respectively, for the eight different energies and
for 30\%-40\% mid-central collisions. The kaon correlations are shown in
Fig. \ref{plot_R2_K} at 200 GeV and 30\%-40\% centrality. The kaon
correlations at lower energies are similar, but become increasingly noisy
due to the weakening production of kaons (and the fewer number of
experimental events) as the energy is decreased. 

\begin{figure*}[htb] 
    \subfloat[pions]{\includegraphics[trim= 0cm 0cm 0cm 0cm,clip,scale=0.45]{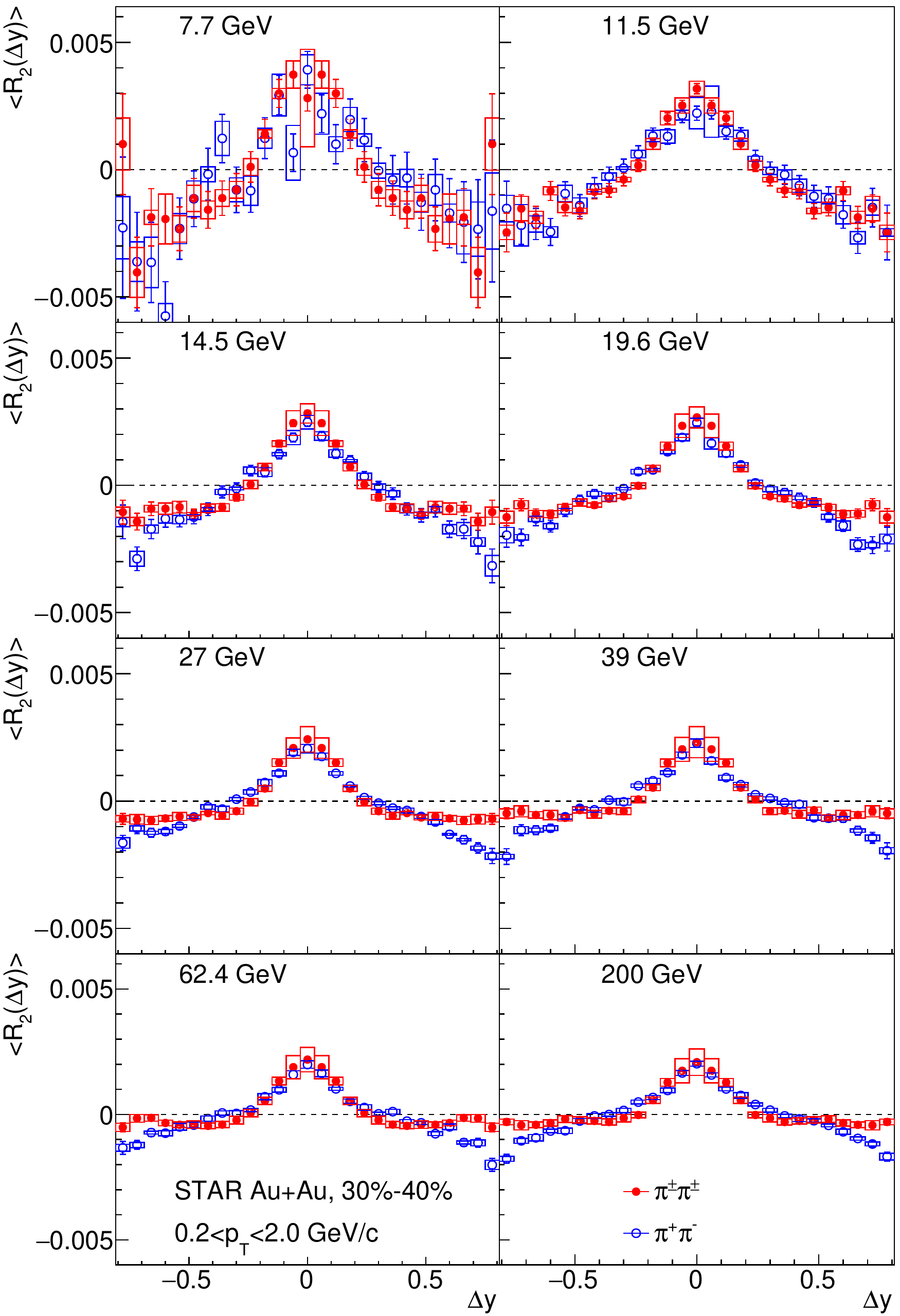} }
    \qquad \qquad
    \subfloat[protons]{\includegraphics[trim= 0cm 0cm 0cm 0cm,clip,scale=0.45]{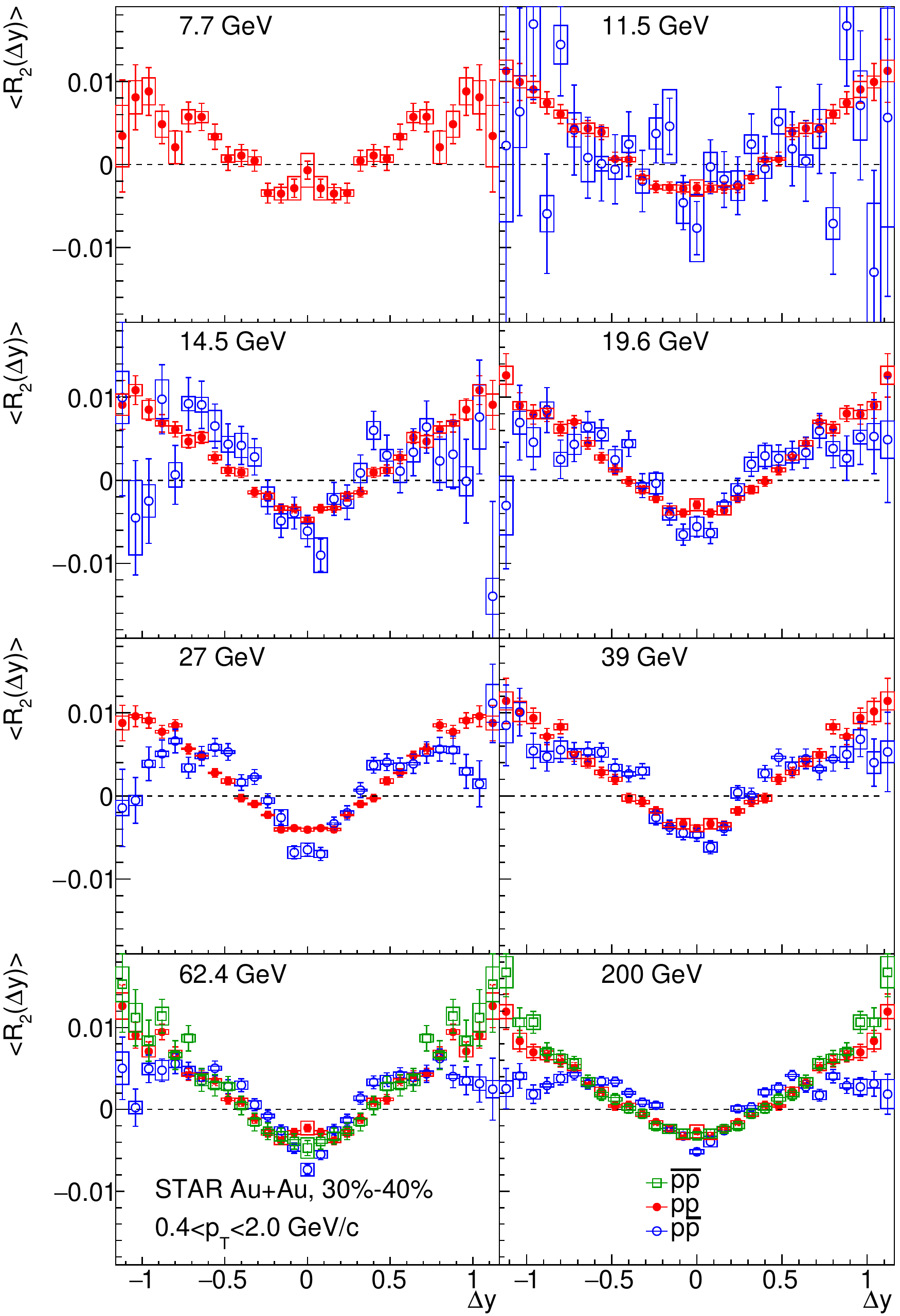} }
    \caption{\small Projection of correlation function $\langle R_{2}(\Delta y)\rangle$ of like-sign (red) and unlike-sign 
(blue) pions (left) and protons (right) in Au+Au collisions at 30\%-40\% centrality and eight 
different energies from 7.7 GeV (top left) to 200 GeV (bottom right). Also shown at the highest 
beam energies in the right frames are the antiproton-antiproton correlations.}
 \label{plot_R2dy}
\end{figure*} 

The like-sign correlations for pions and kaons are the average of the
like-sign positive and like-sign negative correlation functions. For
protons, the like-sign positive and like-sign negative are separately
studied. The like-sign antiproton correlation functions are statistically
significant only at the highest beam energies. 

The correlation functions shown in Figs. \ref{plot_R2_pi}-\ref{plot_R2_K} 
reflect the different physical mechanisms occurring in Au+Au collisions at 30\%-40\% centrality. 
Energy-momentum conservation and dijet
fragmentation generally contribute to produce the away-side ridge at
$\Delta\varphi\sim180^{\circ}$, and collective elliptic flow is
responsible for the double ridge structure at $\Delta\varphi=0^{\circ}$
and $180^{\circ}$. These general features depend weakly on the beam
energy for both the like-sign and unlike-sign charge combinations. The
correlations of pions and kaons exhibit a peak at
$(\Delta y,\Delta\varphi)\sim0$ that would typically be associated
with the short-range mechanisms of minijet string breaking, femtoscopic
correlations, and resonance decay. Femtoscopic correlations include
quantum-statistical effects, Coulomb, and strong interactions and can be
positive or negative.

\begin{figure*}[htb] 
 \subfloat[pions]{\includegraphics[trim= 0cm 0cm 0cm 0cm,clip,scale=0.41]{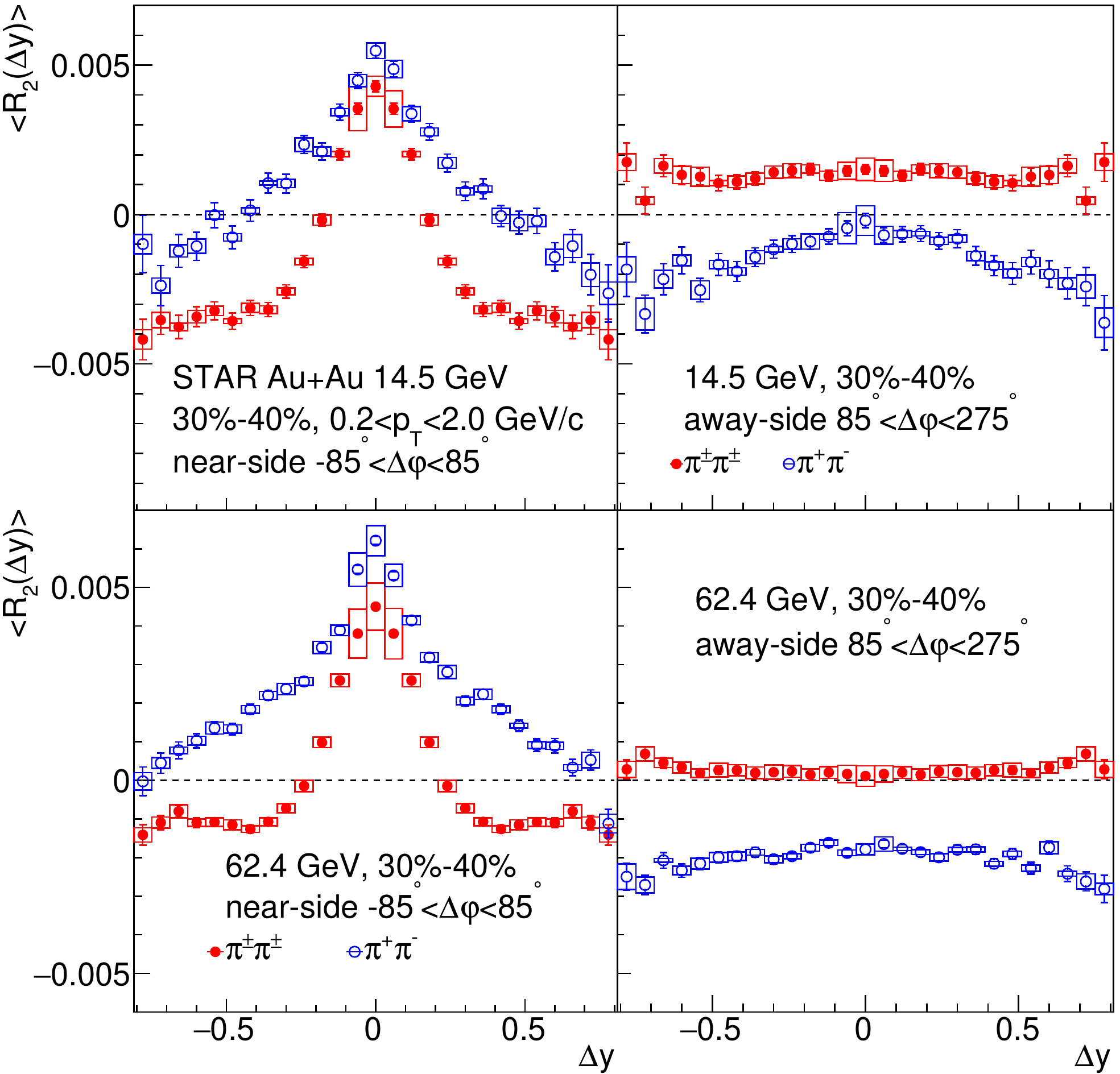} }
    \qquad  \qquad
   \subfloat[protons]{\includegraphics[trim= 0cm 0cm 0cm 0cm,clip,scale=0.41]{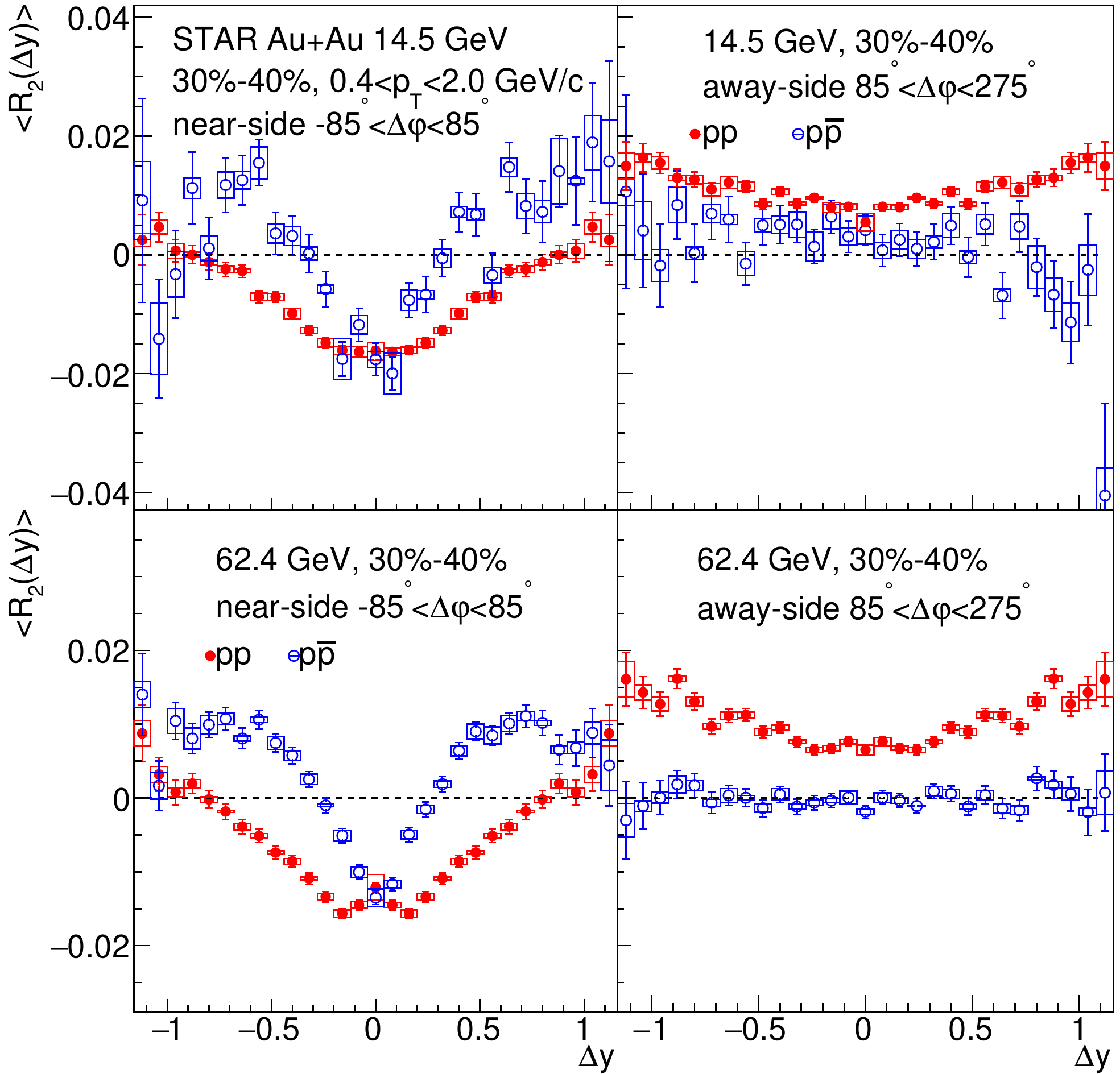} }
   \caption{\small Near-side and away-side $\langle R_{2}(\Delta y)\rangle$ projection of 
like-sign (red) and unlike-sign (blue) pions (left) and protons (right) in Au+Au collisions 
at 14.5 GeV (top) and 62.4 GeV (bottom), 30\%-40\% centrality.}   
 \label{plot_R2dyssas}
\end{figure*}

The strong near side peaks in the like-sign two-pion correlations shown
in Fig. \ref{plot_R2_pi} ($p_{T}<2$ GeV/$c$) are predominantly
femtoscopic in nature. These peaks can be cleanly excised by removing the
(very small) fraction of pairs with $\Delta q<100$ MeV/$c$, where $\Delta q$
is the modulus of the energy-momentum four-vector difference of the two particles in each
pair. Such a cut has very little effect on the unlike-sign pion
correlations because quantum-statistical effects do not occur for
distinguishable particles.

\begin{figure*}[htb] 
 \subfloat[pions]{\includegraphics[trim= 0cm 0cm 0cm 0cm,clip,scale=0.41]{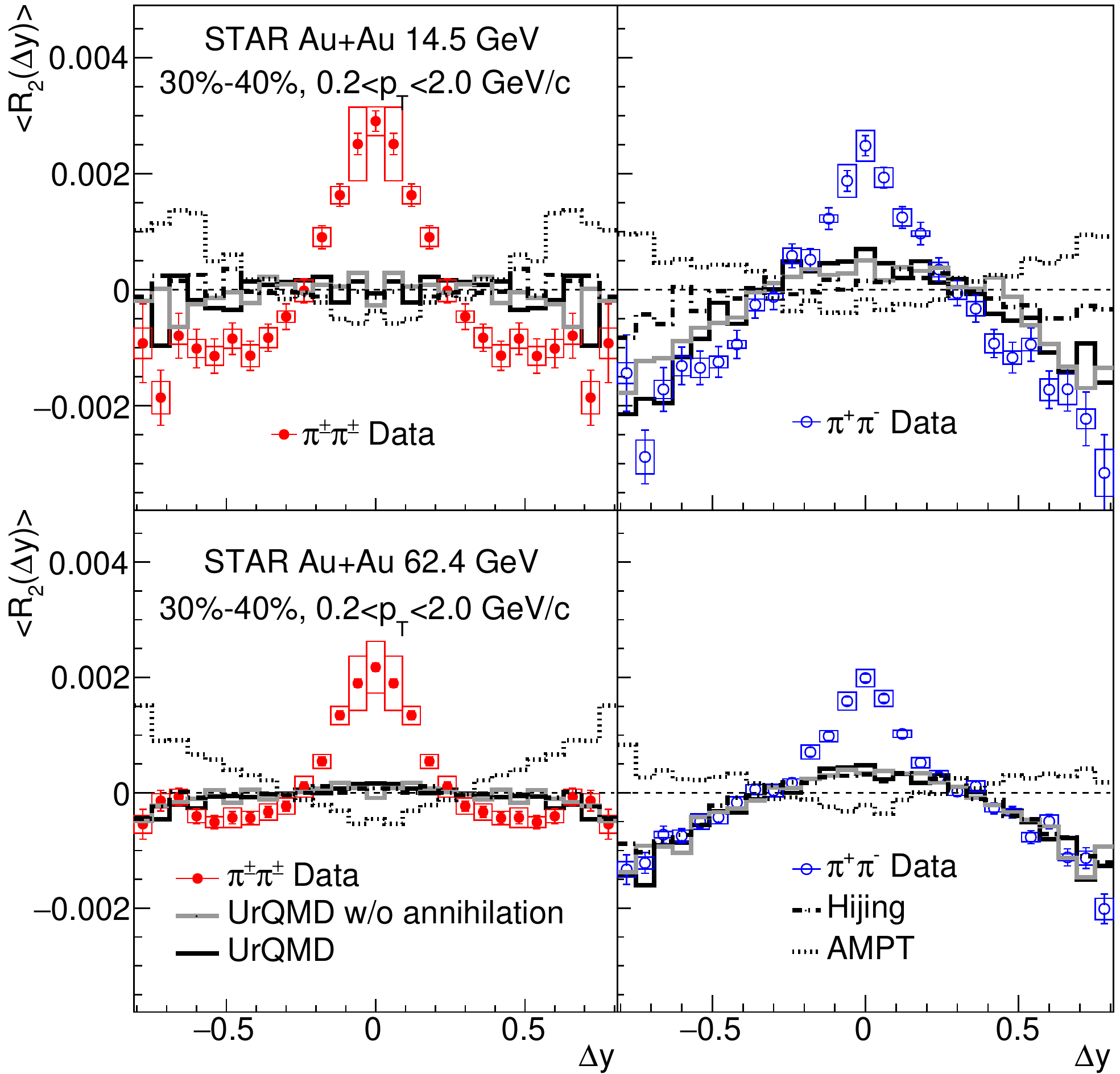} }
    \qquad \qquad 
   \subfloat[protons]{\includegraphics[trim= 0cm 0cm 0cm 0cm,clip,scale=0.41]{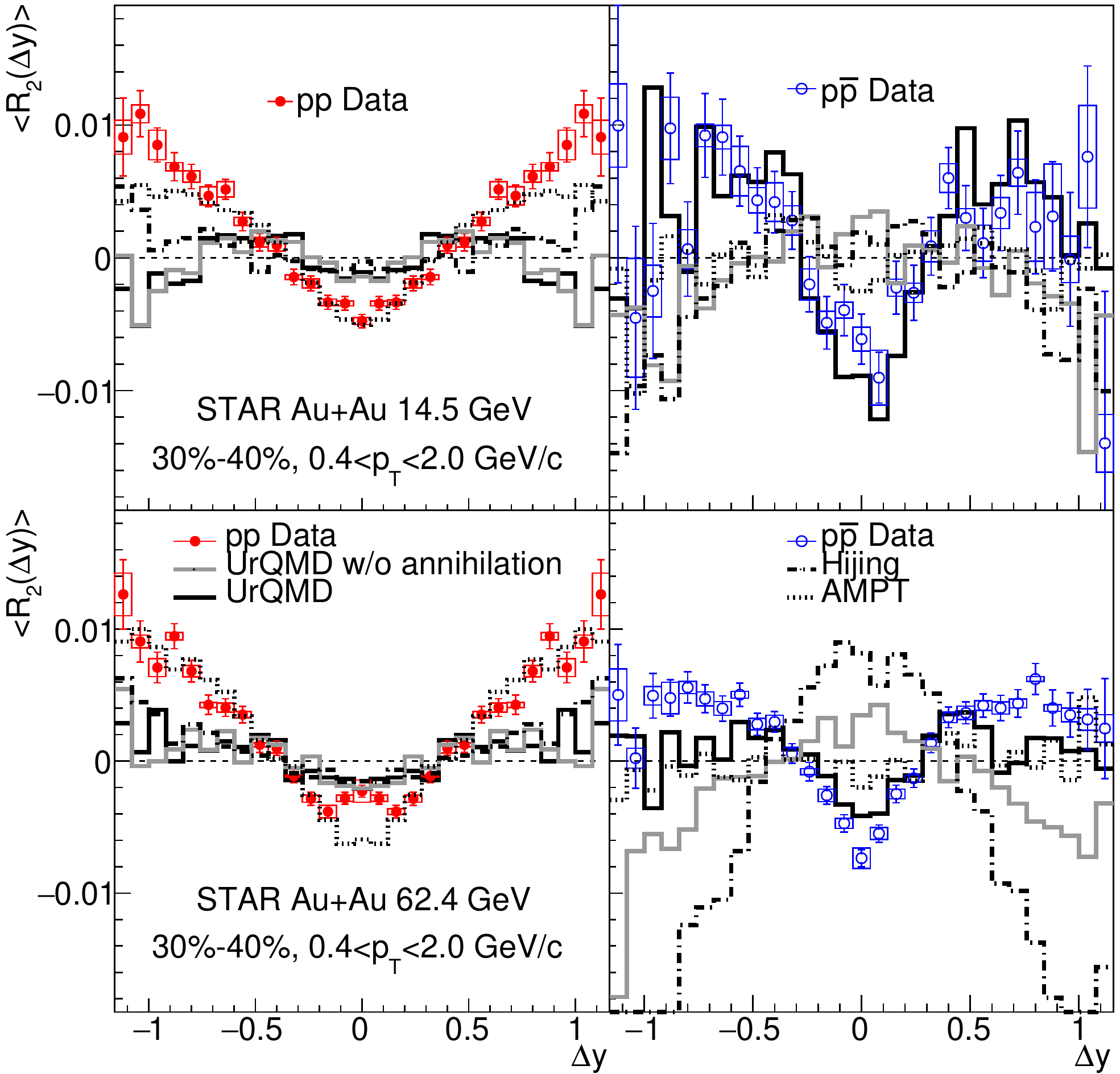} }
   \caption{\small Projection of correlation function $\langle R_{2}(\Delta y)\rangle$ of like-sign (red) and unlike-sign 
(blue) pions (left) and protons (right) in Au+Au collisions at 14.5 GeV (top) and 62.4 GeV 
(bottom), 30\%-40\% centrality compared with the UrQMD (solid line), Hijing (dash-dotted line), 
and AMPT (dotted line) simulations.  }   
 \label{plot_R2dy_model}
\end{figure*}

The near-side peak in the unlike-sign kaon correlations is wider in
$(\Delta y,\Delta\varphi)$ compared to the like-sign kaons in Fig.
\ref{plot_R2_K}. This near-side correlation in unlike-sign kaons is in
the shape of a caldera centered at $(\Delta y,\Delta\varphi)\sim0$ 
which results from K$^{+}$K$^{-}$ pairs that are the daughters of
$\phi$(1020) mesons \cite{pdg2018,Abelev2009}.

The proton correlation functions are qualitatively similar to those for
pions and kaons on the away side in $\Delta\varphi$. However, a
significant difference is observed on the near side,
$(\Delta y,\Delta\varphi)\sim0$. The values of the like-sign proton
correlation functions show a wide suppression on the near side. Upon this
wide anticorrelation may sit a narrow peak at $(\Delta
y,\Delta\varphi)\sim0$.

For the unlike-sign proton pairs, a prominent near-side ridge along the
$\Delta y$ axis is observed for the larger values of $\Delta y$.  At
smaller values of $\Delta y$, a clear anticorrelation with respect to this
ridge is observed. This anticorrelation in unlike-sign proton pairs near
$(\Delta y,\Delta\varphi)\sim0$ is narrower in $\Delta y$ than the
near-side anticorrelation observed for the like-sign proton pairs.

The projections of the angular correlation functions onto the $\Delta y$ axis (integrated over
all azimuthal angles) for like-sign and unlike-sign pion and proton pairs in 30\%-40\% central 
collisions are shown in Fig.
\ref{plot_R2dy}. The proton pair correlations and pion pair correlations differ significantly
at all eight energies and for both like-sign and unlike-sign combinations. The 
pion correlations show an enhancement around $\Delta y\sim0$ 
which decreases slightly with increasing beam energy.  

In contrast, both the like-sign and unlike-sign proton correlations show an anticorrelation
near $\Delta y\sim0$ at all eight energies. These anticorrelations are
remarkably weakly-dependent on the beam energy. 
The values of the correlation functions near $\Delta y\sim0$ for 
the like-sign (red) and unlike-sign (blue) pairs are comparable at all
eight energies. 
At larger values of the rapidity difference, the like-sign
proton correlations continue to rise roughly linearly, while the values
for unlike-sign pairs level off to form the near-side ridge seen in Fig. \ref{plot_R2_p}.

Also shown on the lower right in this figure are the like-sign antiproton correlation functions 
(green) at the two highest beam energies. Lower beam energies result in considerably fewer
antiprotons, and thus much more uncertain correlation functions, so the like-sign 
antiproton results are not shown for clarity. The like-sign antiproton
correlations are consistent with those for like-sign protons.

The projection of $R_{2}(\Delta y,\Delta\varphi)$ into $\Delta y$, 
averaged over $|\Delta\varphi|<85^{\circ}$ (a ``near-side projection") or averaged
over $85^{\circ}\le|\Delta\varphi|\le275^{\circ}$ (an ``away-side projection") is
shown in Fig. \ref{plot_R2dyssas} for the like-sign and unlike-sign pion and proton pairs
at 14.5 and 62.4 GeV in 30\%-40\% central collisions. 
The away-side projections of the pion and proton correlations are roughly
flat versus the rapidity difference as seen in the two right frames
of Figs. \ref{plot_R2dyssas}a and \ref{plot_R2dyssas}b.
There is a slight suppression on the away-side for the like-sign protons 
due to the wider near-side anticorrelation in $(\Delta y,\Delta\varphi)$ 
(compared to that for the the unlike-sign pairs) which was shown
in Fig. \ref{plot_R2_p}. 
The correlations of the like-sign pions and protons (red) are larger than those 
for the unlike-sign pairs (blue) on the away-side.

The $\Delta y$ dependence of the correlations on the near-side 
explored in Fig. \ref{plot_R2dy} come into better focus when requiring 
each pair is also on the near-side azimuthally, and are shown in Fig. \ref{plot_R2dyssas}. 
Here, the correlations of the unlike-sign pions is larger than those for like-sign pairs,
which is opposite to the behavior observed on the away-side. 
The near-side proton correlations shown in Fig. \ref{plot_R2dyssas}b indicate 
an anticorrelation in both the like-sign and unlike-sign charge combinations.
Here it is again seen, as in Fig. \ref{plot_R2_p}, that the unlike-sign proton
anticorrelation is much narrower in $\Delta y$ compared to that for the 
like-sign proton pairs.

The unlike-sign pion correlations shown in Fig. \ref{plot_R2dyssas}a are much wider
on the near side (left frames) in $\Delta y$ than those for the like-sign 
pion pairs. This is presumably due to local charge
conservation in unlike-sign pairs \cite{Bozek2012}. 
The effects of local charge conservation on the proton correlations is 
less clear, but the difference of the unlike-sign and like-sign correlation
functions are similar for both pions and protons at the larger values 
of $\Delta y$. Therefore, local charge conservation may contribute to
the faster rise in the unlike-sign proton correlations compared to the 
like-sign pairs.

The measured pion and proton correlation functions were compared to those 
obtained using the events generated by several 
model event generators. The analysis was done for simulated events using 
UrQMD v3.4 \cite{Bass1998}, Hijing v1.411 \cite{Wang1991}, and AMPT v2.26t7b \cite{Lin2005}. 
The UrQMD model is based on the covariant propagation of color
strings, constituent quarks, and diquarks accompanied by mesonic and baryonic degrees
of freedom. It simulates multiple interactions of ingoing and newly produced
particles, the excitation and fragmentation of color strings, and the formation and
decay of hadronic resonances \cite{Bass1998}. The Hijing model is used to
study jet and multiparticle production in high energy p+p, p+A, and A+A collisions at
RHIC and LHC energies. The model includes multiple minijet production, nuclear
shadowing of the parton distribution functions, and a schematic mechanism of jet
interactions in dense matter, which contains many sources of long and short-range
correlations \cite{Wang1991}. A ``multi-phase transport model," (AMPT) uses the
Hijing model for generating the initial conditions, then models the partonic scattering, 
string fragmentation using the Lund model, hadronization via quark coalescence, 
and finally hadronic rescattering \cite{Lin2005}. 

Approximately 30M minimum bias events were generated
using the default parameters for each model. Additional model data sets of the 
same significance were also generated following the modification of 
specific model parameters in order to further explore specific topics. 
The centrality of the model events was determined by integrating the minimum 
bias distributions of the charged particle multiplicities calculated with the same 
kinematic cuts as were used for the analysis of the experimental data. 

Figure \ref{plot_R2dy_model} depicts the comparison of the experimental
and model results for like-sign and unlike-sign pions and protons at 14.5
GeV and 62.4 GeV in 30\%-40\% mid-central collisions. None of the three
models describes the observed pion correlations at small values of the
rapidity difference, $\Delta y$. As described above, this strong short
range peak in the like sign correlations appears to be predominantly
femtoscopic in origin as it can be removed by removing pairs with
$\Delta q<100$ MeV/$c$. This can be expected as the models generally 
make no attempt to describe femtoscopy in their default configurations.
However, the disagreement between the data and models for the unlike-sign
pion short-range correlations cannot be explained by femtoscopy as the
same $\Delta q$ cut does not remove the short-range correlation, and the
particles in the pair are distinguishable. 

\begin{figure*}[htb] 
    \subfloat[pions]{\includegraphics[trim= 0cm 0cm 0cm 0cm,clip,scale=0.41]{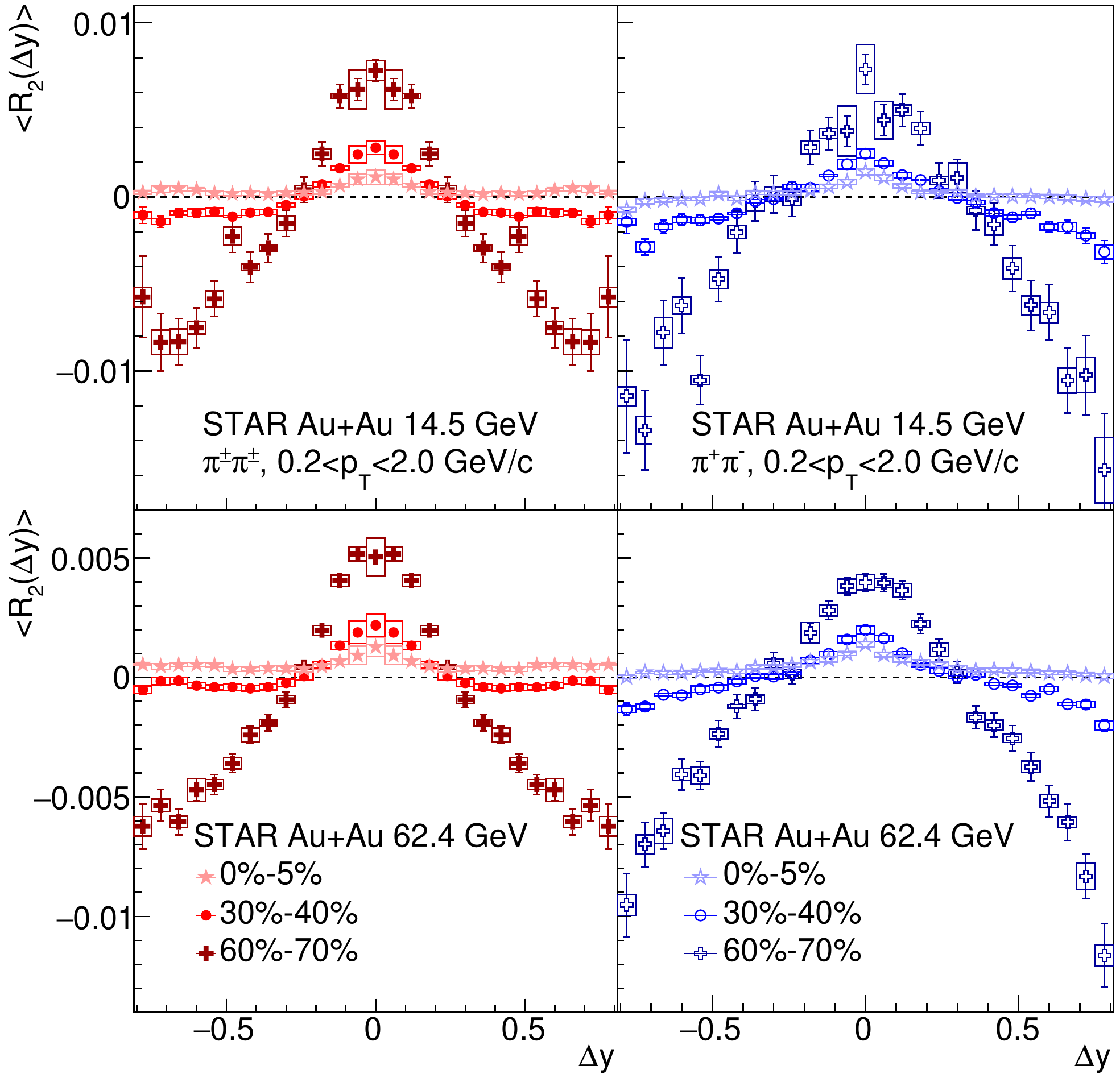} }
    \qquad  \qquad
    \subfloat[protons]{\includegraphics[trim= 0cm 0cm 0cm 0cm,clip,scale=0.41]{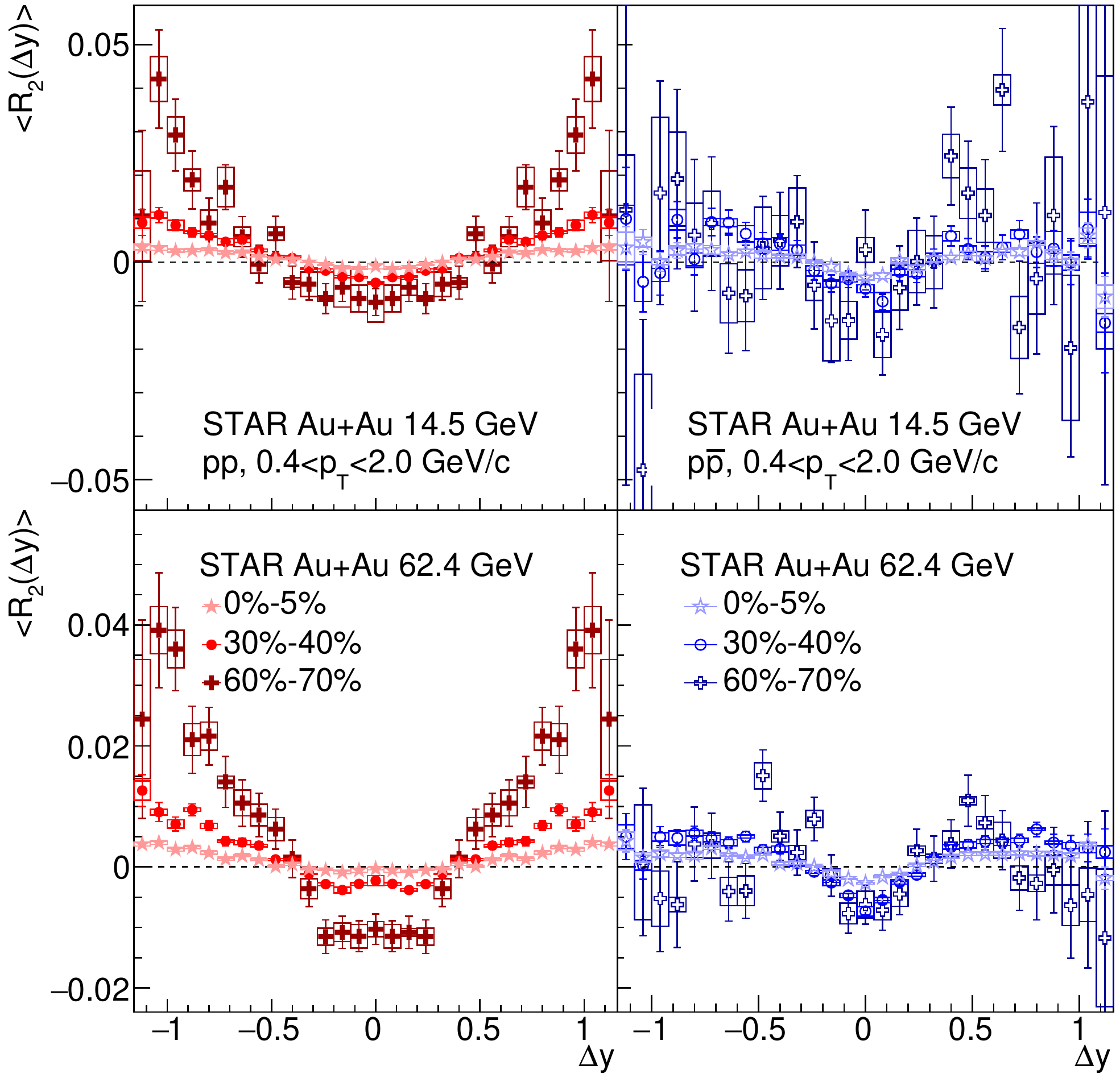} }   
    \caption{\small Projection of correlation function $\langle R_{2}(\Delta y)\rangle$ 
of like-sign (red) and unlike-sign (blue) pions (left) and protons (right) in Au+Au collisions 
at 14.5 GeV (top) and 62.4 GeV (bottom) for the most central 0\%-5\%, mid-central 30\%-40\% and 
peripheral 60\%-70\% events.}
 \label{plot_R2dy_cent}
\end{figure*}

\begin{figure*}[htb] 
    \subfloat[pions]{\includegraphics[trim= 0cm 0cm 0cm 0cm,clip,scale=0.41]{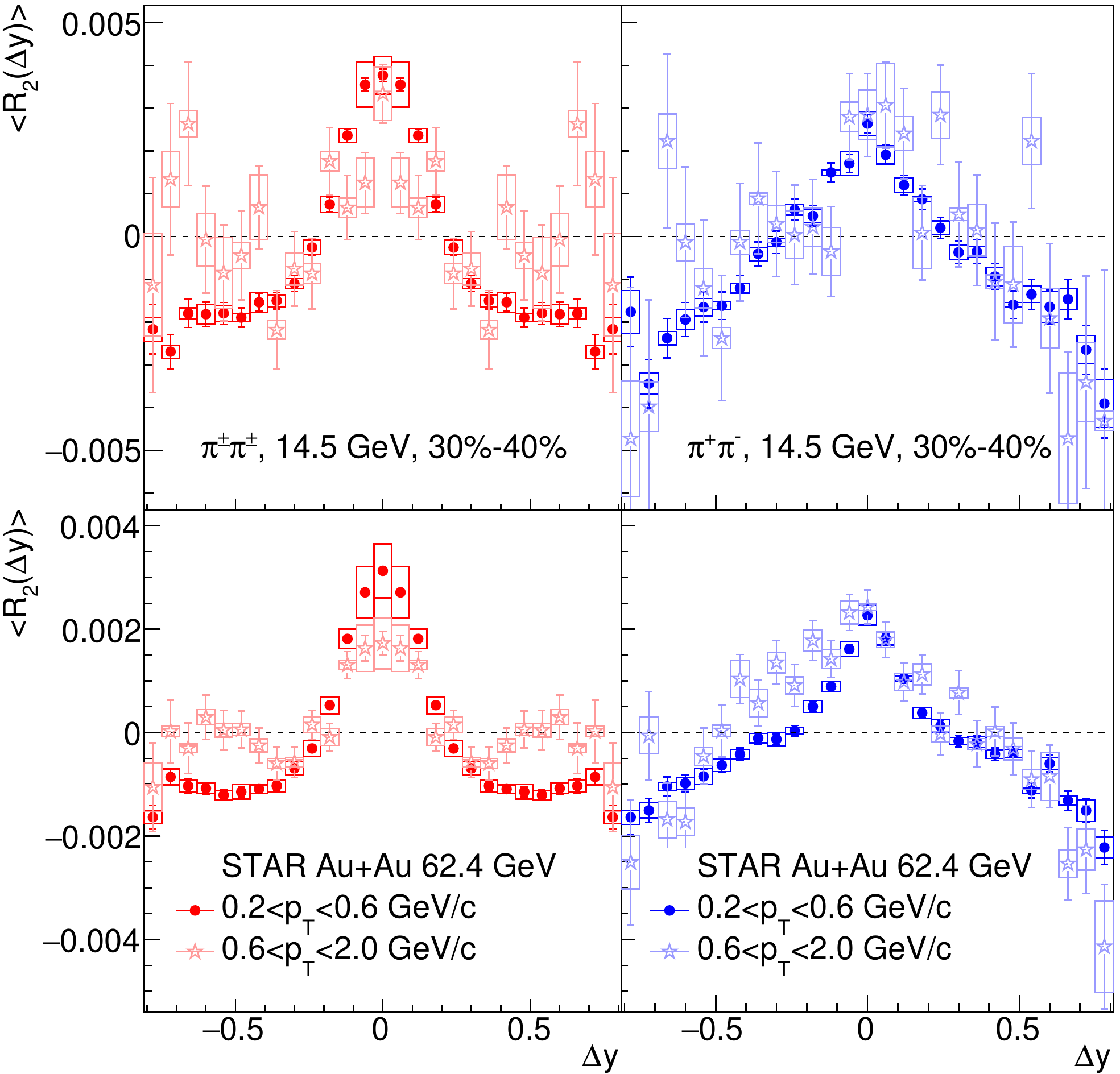} }
    \qquad
    \subfloat[protons]{\includegraphics[trim= 0cm 0cm 0cm 0cm,clip,scale=0.41]{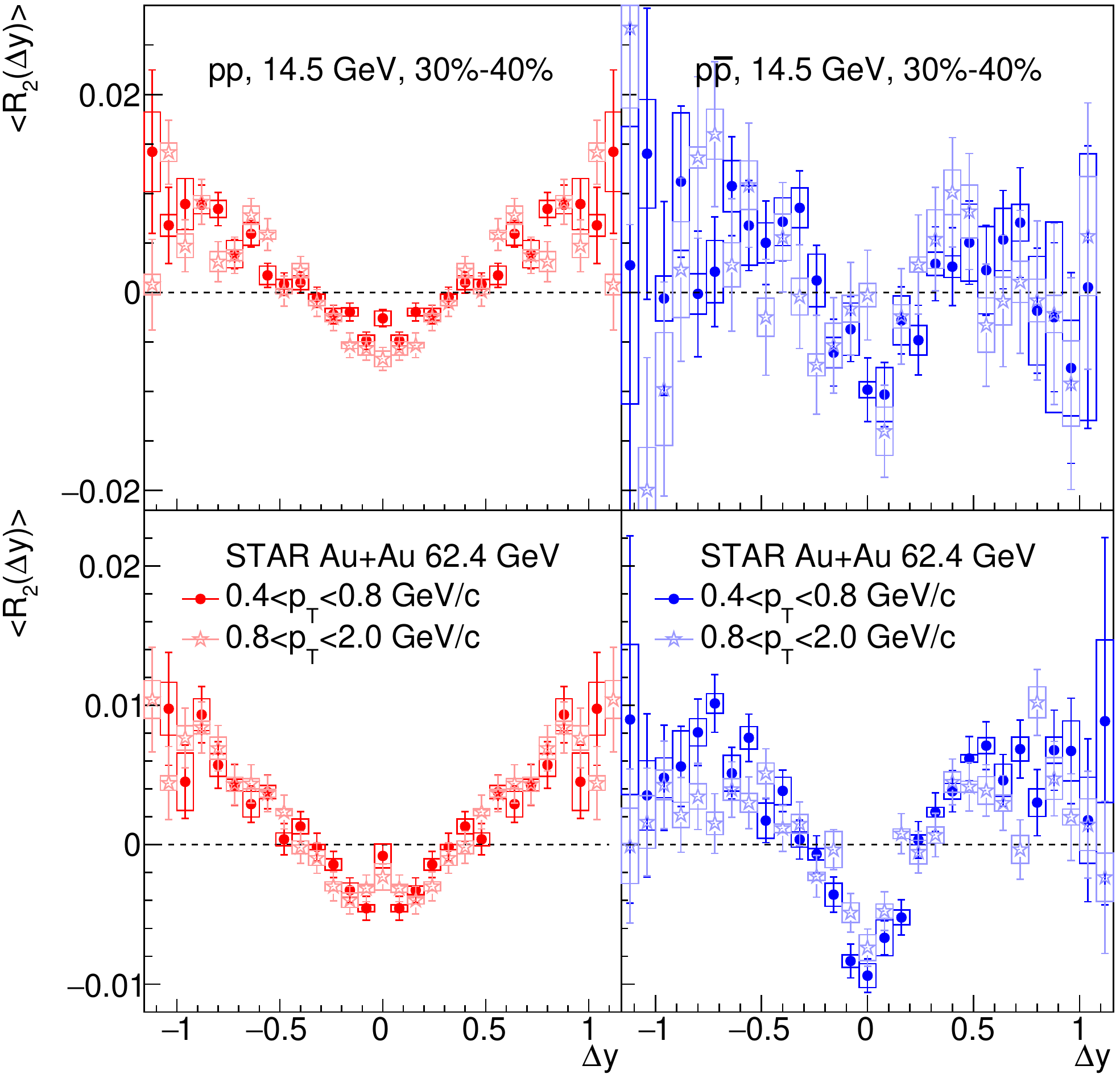} }
    \caption{\small Projection of correlation function $\langle R_{2}(\Delta y)\rangle$ of like-sign 
(red) and unlike-sign (blue) pions (left) and protons (right) in low and high $p_{T}$ in 
Au+Au collisions at 14.5 GeV (top) and 62.4 GeV (bottom) in 30\%-40\% centrality. }
\label{plot_R2dy_pT}
\end{figure*} 

The UrQMD and Hijing models were more successful than AMPT in reproducing
the correlations of unlike-sign pions at larger values of $\Delta y$.
This may be the result of a stricter local charge conservation in UrQMD
and Hijing compared to AMPT \cite{Pan2014}.

For the proton correlations, Hijing does not describe the data, while
UrQMD and AMPT qualitatively predict a small suppression near
$\Delta y\sim0$ of like-sign and unlike-sign protons, respectively,
but do not reproduce the observed correlations at larger values of
$\Delta y$. The AMPT model can reproduce the observed anticorrelations
for like-sign protons (but fails for unlike-sign protons), while the
UrQMD model can describe the unlike-sign protons (but fails for like-sign
protons).

Also shown in Fig. \ref{plot_R2dy_model} are the results from UrQMD when
baryon annihilation is turned off via a user parameter\footnote{UrQMD ``CTOption(19)" was changed from zero to one.  }. 
The unlike-sign proton
correlations in these events now no longer reproduce those seen in the data near 
$\Delta y\sim0$, and in fact they look quite similar to those obtained from
Hijing and AMPT. This suggests that the anticorrelation in unlike-sign proton pairs
on the near side in $\Delta\varphi$ and at short range in $\Delta y$, best seen in the 
right frames of Fig. \ref{plot_R2_p}, results from baryon-antibaryon annihilation. 

The anticorrelation in like-sign protons is broader and longer range. Similar
two-proton anticorrelations (see also Ref. \cite{prabhat}) were reported in the small collision system
of e$^{+}$+e$^{-}$ annihilation at $\sqrt{s}=29$ GeV by the TPC/Two-Gamma
Collaboration \cite{Twogamma}, and in p+p collisions at $\sqrt{s}=7$ TeV
by the ALICE Collaboration \cite{ALICEproton}. We report this observation
here for the first time in the large collision system of Au+Au.
Although there is a qualitative similarity in the (anti)correlations of
like-sign protons between the small and large systems, there is no such
agreement for unlike-sign protons.

The observed proton anticorrelations in e$^{+}$+e$^{-}$ annihilation at
$\sqrt{s}=29$ GeV were suggested \cite{Twogamma} to result from local
baryon number conservation during the hadronization process and the
``energy cost" required to produce two baryons during the fragmentation
of a single string. According to the string hadronization model
\cite{Lund}, two baryons produced in a single fragmentation should be
separated by at least one particle with a different baryon number
\cite{Twogamma,ALICEproton}. Furthermore, the probability of producing
two baryons in a single fragmentation in low energies is suppressed,
since a minimum of two baryons and two antibaryons would be required to
produce two like-sign baryons while conserving baryon number. This
explanation could be reasonable at the low beam energy of 29 GeV.
However, such an energy constraint seems unlikely in the case of the p+p
collisions at $\sqrt{s}=7$ TeV measured by ALICE, which showed a similar near-side
suppression. In the ALICE study \cite{ALICEproton}, the possibility that
the like-sign proton correlations were suppressed on the near-side by
Fermi-Dirac statistics was ruled out as the p$\Lambda$+$\bar{\rm
p}\bar{\Lambda}$ correlators also showed the same anticorrelations. Other
ideas like the effects of the momentum transfer during the interaction,
Coulomb repulsion, local baryon number conservation, and energy
conservation were also discussed in Ref. \cite{ALICEproton}, but none of
these were seen as entirely successful in explaining their observed baryon
anticorrelations.

The pion and
proton correlations were studied in different centralities from the most central to the most peripheral
collisions. The results of the most central 0\%-5\%, mid-central 30\%-40\%, and
peripheral 60\%-70\% events in Au+Au collisions at the low energy of 14.5 GeV, and the higher
energy of 62.4 GeV, are shown in Fig. \ref{plot_R2dy_cent}. A strong centrality dependence is
observed in the pion and proton correlations. In both cases, the 
(anti)correlations decrease, {\it i.e.}, $R_2$ approaches zero from above or below, as the 
collisions become more central. This is consistent with the usual picture of the dilution
of the correlations due to the increasing number of particle sources as the collisions
become more central.

These correlations were also studied in two different transverse
momentum ranges. The low-$p_{T}$ range for pions and
protons was 0.2-0.6 GeV/$c$ and 0.4-0.8 GeV/$c$, respectively, while the
high-$p_{T}$ range for pions and protons was 0.6-2.0 GeV/$c$ and 0.8-2.0
GeV/$c$, respectively. In Fig. \ref{plot_R2dy_pT}, the pion and proton correlations in these two $p_{T}$
ranges are shown for 30\%-40\% mid-central collisions
at 14.5 GeV and 62.4 GeV. The proton correlations show no significant dependence on the transverse momentum
range for both the unlike- and like-sign charge combinations. There is a 
more significant $p_{T}$ dependence for the like-sign pion correlations
at large $\Delta y$, while the unlike-sign pions do not show a significant $p_{T}$ 
dependence.

The influence of femtoscopic correlations on the
observed proton anticorrelations was also studied. A relative invariant momentum cut
was set based on the values of the effective source size measured by STAR \cite{pHBT, piHBT}.
This cut would be expected to suppress all femtoscopic contributions. The bins
of the correlation function affected by such a cut is limited to the rather small region of $\Delta y<0.1$. 
This is much narrower than the observed width of the observed proton anticorrelations.

\section{Summary and Conclusions\label{sec:summary}}

The two-particle angular correlation functions were studied for like-sign
and unlike-sign pion, kaon, and proton pairs in the Beam Energy Scan data
collected by the STAR experiment. The energy, centrality, and $p_{T}$
dependence of the correlations were investigated. No nonmonotonic
behavior was observed in any of the two-particle angular correlation
functions as a function of the beam energy from 7.7 to 200 GeV and indeed
the dependence on the beam energy is quite weak overall. The experimental
results were also compared to those obtained from the models UrQMD,
Hijing, and AMPT. 

The expected near-side peak was observed in the pion and kaon
correlations which is associated with short-range mechanisms. In the case
of the like-sign two-pion correlations, this peak appears to be
predominantly femtoscopic in the kinematic range of this analysis as it
can be removed by removing pairs with a relative four-vector
difference of less than 100 MeV/$c$. The amplitudes of the correlations
decrease with increasing beam energy and decrease as the collisions
become more central, and are at most weakly dependent on the transverse
momentum in two wide bins of this variable. A strong near-side
ring-shaped positive correlation was observed in the unlike-sign kaon
correlations resulting from the strongly correlated pairs from
$\phi$(1020) decays.

In contrast to the meson correlations, the proton pairs exhibit a
significant near-side anticorrelation at all beam energies. This proton
anticorrelation has already been observed in small systems and is here
reported for the first time in the large collision system of Au+Au.
This anticorrelation was observed in both like-sign and unlike-sign
(anti)proton pairs, and it is wider in relative rapidity, $\Delta y$,
for the like-sign charge combination as compared to the unlike-sign
combination. The model comparisons imply that the anticorrelation in the
unlike-sign proton pairs is caused by baryon-antibaryon
annihilation. A description of the cause of the stronger and
longer-range anticorrelation in the like-sign proton pairs is not yet in
hand. This like-sign proton anticorrelation is apparently $p_{T}$-independent, 
decreasing with increasing beam energy, and decreasing
as the collisions become more central. 

%
%
%
\begin{acknowledgments}
We thank the RHIC Operations Group and RCF at BNL, the NERSC Center at LBNL, and the Open Science Grid consortium for providing resources and support.  This work was supported in part by the Office of Nuclear Physics within the U.S. DOE Office of Science, the U.S. National Science Foundation, the Ministry of Education and Science of the Russian Federation, National Natural Science Foundation of China, Chinese Academy of Science, the Ministry of Science and Technology of China and the Chinese Ministry of Education, the National Research Foundation of Korea, Czech Science Foundation and Ministry of Education, Youth and Sports of the Czech Republic, Hungarian National Research, Development and Innovation Office, New National Excellency Programme of the Hungarian Ministry of Human Capacities, Department of Atomic Energy and Department of Science and Technology of the Government of India, the National Science Centre of Poland, the Ministry  of Science, Education and Sports of the Republic of Croatia, RosAtom of Russia and German Bundesministerium fur Bildung, Wissenschaft, Forschung and Technologie (BMBF) and the Helmholtz Association.\end{acknowledgments}

\bibliography{References}

\end{document}